\documentclass{article}

\usepackage{arxiv}

\usepackage[utf8]{inputenc} 
\usepackage[T1]{fontenc}    
\usepackage{hyperref}       
\usepackage{url}            
\usepackage{booktabs}       
\usepackage{amsfonts}       
\usepackage{nicefrac}       
\usepackage{microtype}      
\usepackage{xcolor}         
\usepackage{amsfonts,mathtools,mathrsfs,mathtools,color}

\usepackage{amssymb, amsmath, amsthm}

\usepackage{algorithm}
\usepackage{algpseudocode}

\newtheorem{theorem}{Theorem}
\newtheorem{theorem*}{Theorem}

\newcommand{\argmax}{\text{argmax}}

\title{Polynomial Expectation Property for Max-Polymatrix Games}

\author{
 Howard Dai \\
  Department of Computer Science\\
  Yale University\\
  New Haven, CT 06511 \\
  \texttt{howard.dai@yale.edu} \\
}

\begin{document}
\maketitle
\begin{abstract}
  We address an open question proposed by Papadimitriou and Roughgarden \cite{papadimitriou2008computing} on the computability of correlated equilibria in a variant of polymatrix where each player's utility is the maximum of their edge payoffs. We demonstrate that this max-variant game has the polynomial expectation property, and the results of \cite{papadimitriou2008computing} can thus be applied. We propose ideas for extending these findings to other variants of polymatrix games, as well as briefly address the broader question of necessity for the polynomial expectation property when computing correlated equilibria. 
\end{abstract}



\section{Background}
Our work revolves around the findings of \cite{papadimitriou2008computing}, which demonstrates that being able to efficiently compute the expected value of a player's utility is a sufficient condition for being able to efficiently compute a correlated equilibrium. To contextualize our work, we briefly cover definitions and the main findings from this paper in the following subsections. 

\subsection{Succinct Games}
For consistency, we use the same notation in \cite{papadimitriou2008computing}. In particular, a \textit{succinct game} $G = (I, T, U)$ consists of efficiently recognizable inputs $I$, a polynomial-time algorithm $T$ to represent the \textit{type}, and a polynomial-time algorithm $U$ to compute utility. In particular, given an input $z \in I$, $T(z)$ returns $n, (t_1, \dots, t_n)$, which correspond to the number of players and cardinalities of each players' strategy sets, respectively. If $n$ and $m = \max_p{|t_p|}$ are polynomially bounded in $|z|$, the game is of \textit{polynomial type}. $U(z,p,s)$ returns the utility player $p$ receives given strategies $s = (s_1, s_2, \dots s_n)$. 

\subsection{Correlated Equilibria}
A \textit{correlated equilibrium} is a distribution $x$ over a set of strategy profiles $S$, with the condition that for every player $p$ and strategy $i$, player $p$ has no incentive to swap to a different strategy $j$. In particular, this can be written as:
$$\mathbb{E}_{s \sim x \, | s_p = i} [U(z,p,(i, s_{-p}))] \geq \mathbb{E}_{s \sim x \, | s_p = i} [U(z,p,(j, s_{-p}))] \, \, \, \forall j \in S_p $$
These conditions can be written as a linear program, with one row per choice of player and two strategies (approximately $m^2n$ rows), such that the linear program is unbounded if and only if a correlated equilibrium exists. As shown in \cite{papadimitriou2008computing}, running the ellipsoid algorithm on the infeasible dual (``Ellipsoid Against Hope") allows for a solution to the program (and thus a correlated equilibrium) to be expressed as a convex combination of polynomially many product distributions generated by the individual violated constraints at each step. 

All steps in this process can be computed in polynomial time, except for one: the multiplication of the constraint matrix by the matrix containing the product distributions, which requires the computation of inner products of size $m^n$. Each inner product, when expanded, can be expressed as a difference of two expected utilities; thus, the product can be computed in polynomial time if the expected utility of a player can be computed in polynomial time. This is where the main result of the paper comes from: being able to compute the expected utility is sufficient to construct a correlated equilibrium in polynomial time. 

\subsection{Polynomial Expectation Property} \label{polyexp}
A succinct game $G$ has the polynomial expectation property if there exists a polynomial-time algorithm $\mathcal{E}$ which, given a \textit{product distribution} $x = \prod_{i=1}^n x_i$, returns a player $p$'s expected utility $u^p$ under this distribution:
$$\mathcal{E}(z, p, x) = \mathbb{E}_{s \sim x} [U(z, p, s)]$$
Note that the most naive algorithm would compute expected utility as such: 
$$\mathcal{E}(z, p, x) = \sum_{s\in S} x(s)U(z, p, s)$$
This would require a number of calls to $U$ equivalent to the number of strategy profiles in $S$, which is on the order of $m^n$. However, many well-known classes of games, such as polymatrix games, allow for a polynomial time computation of this expression. 

The main result of \cite{papadimitriou2008computing} is as follows:

\begin{theorem} \label{main_thm}
    If $G$ is a succinct game of polynomial type and has the polynomial expectation property, then it has a polynomial correlated equilibrium scheme.
\end{theorem}

For the remainder of this paper, we assume all games are of polynomial type, as this result is irrelevant otherwise. 

\subsection{Polymatrix Games}
In a polymatrix game, input $z$ describes $\binom{n}{2}$ 2-player games between every possible pair of players $(p,q)$, denoted $G_{pq}$, where each player $p$ has the same strategy set $S_p$ in each.

In polymatrix games, the utility function $U$ can be reduced to a series of per-game utility functions $U_{pq}$, such that $U_{pq}(z,p,(s_p,s_q))$ denotes player $p$'s utility in a two-player game against player $q$. In standard polymatrix games, a player $p$'s utility is the sum over all of their individual two-player games:
$$\mathcal{E}(z,p,x) = \mathbb{E}_{s \sim x}[\sum_{q\neq p} U_{pq}(z,p,(s_p,s_q))]$$
By linearity of expectation, we have:
$$\mathcal{E}(z,p,x) = \sum_{q\neq p} \mathbb{E}_{(s_p,s_q) \sim (x_p, x_q)}[U_{pq}(z,p,(s_p,s_q))]$$
$$\mathcal{E}(z,p,x) = \sum_{q\neq p} \sum_{(s_p, s_q)} x_p(s_p) x_q(s_q)U_{pq}(z,p,(s_p,s_q))$$
Thus, we compute $U_{pq}$ on the order of $nm^2$ times. Assuming a polynomial type game, this implies we can compute $\mathcal{E}(z,p,x)$ in polynomial time with respect to $|z|$. 

\section{Polynomial Expectation Property for Max-Variant Polymatrix Games}

In \cite{papadimitriou2008computing}, Papadimitriou and Roughgarden propose an open problem relating to a variant of polymatrix games: 
\begin{quote}
    Our approach works for all kinds of succinct multiplayer games of polynomial type in
the literature known to us. There are, however, artificial examples for which it does
not seem to: A simple one is a variant of polymatrix games in which a player’s utility
is the maximum utility over all games played, as opposed to the sum. The maximum
destroys, of course, linearity of expectation. Is there a correlated equilibrium scheme
for this class?
\end{quote}

We refer to this variation of polymatrix as \textit{Max-Polymatrix}. We display an explicit polynomial-time algorithm for computing expected utility in a Max-Polymatrix game in Algorithm \ref{alg:cap}. We also provide analysis of the algorithm below. On a high level, for a fixed player $p$ and strategy $i$, we can avoid iterating over all possible combinations of opposing player strategies $S_{-p}$ ($\sim m^n$ iterations), by doing casework on which opposing player and strategy provides the maximum utility for player $p$ ($\sim nm$ iterations).  

In Max-Polymatrix, every player $q \neq p$, action $j \in S_q$ corresponds to one potential utility value $u_q(j)$. Then it suffices to find, for each player $q^*$ and action $j$, the probability that $u_{q^*}(j)$ is the maximum over all other $u_{q}(S_{q})$. This is equivalent to the probability that $j$ is drawn, $x_{q^*}(j)$, multiplied by the probability that no other utilities attained against any other $q$ are higher than $u_{q^*}(j)$. This is captured by the variable $c_{q}$, which tracks the probability that $u_q(j) < u_{q^*}(j)$. In particular, if $M = u_{q^*}(j)$ is the current utility value in the loop, we have:
$$c_q = \sum_{k: \, u_q(k) < M } x_q(k) $$
By construction, $M$ is nonincreasing during the loop, so it suffices to subtract $x_q(k)$ from $c_q$ once $u_q(k) = M$ (i.e. $u_q(k)$ becomes the ``current" utility). Note that, in the case of ties, we can assign the same (arbitrary) tiebreaking rule to the algorithm's ``argmax" and to the ``$>$" operator, and this invariant will  remain true. 

For each fixed action $i$, the loop iterates over every other player, action pair, performing $O(n)$ operations for each iteration, giving $O(n^2m)$ runtime for the inner loop. In addition, sorting $n$ lists of length $m$ to the inner loop can be done in $O(n m \log m)$ time. This entire process is performed once per action $i$, giving a total of $O(m (n^2m + nm\log m)) = O(n^2m^2 \log m)$ runtime, which is polynomial in $m, n$. By our assumption of a polynomial-type game, this implies polynomial runtime in $|z|$. Thus, Max-Polymatrix has the polynomial expectation property, and thus has a polynomial correlated equilibrium scheme by \cite{papadimitriou2008computing}. 

\begin{algorithm} 
\caption{Compute Expected Utility in Max-Polymatrix}\label{alg:cap}
\begin{algorithmic}
\State {\bfseries Input:} Input $z$, player $p$, product distribution $x$
\For {action $i$ in $S_p$} 
\State $E(i) \gets 0$ \Comment{exp. utility given fixed action $i$}
\State $c_q \gets 1,\, \, \forall q \neq p$ \Comment{``CDF" variable for each player}
\State $u_q(j) = U_{pq}(z,p,(i,j)) \, \, \forall q \neq p, j \in S_q$ 
\State Sort each $u_q$ in descending order \Comment{Note that $x_q$ should be aligned accordingly as well}
\While {$u \neq \emptyset$}
\State $q^* \gets \argmax_{q\neq p} u_q(0)$ 
\State $E(i) \gets E(i) + u_{q^*}(0) \cdot (x_{q^*}(0) \cdot \prod_{q \neq q^*} c_q)$ 
\State $c_{q^*} \gets c_{q^*} - x_{q^*}(0)$
\State Remove action $0$ from $u_{q^*}$, shift other actions up
\EndWhile 
\EndFor \\ 
\Return $\sum_{i \in S_p} x_p(i) E(i)$ 
\end{algorithmic}
\end{algorithm}

\section{Extensions and Future Directions}

\subsection{Alternative Polymatrix Payoff Functions}
When the polymatrix payoff function is linear, expected utility can be computed trivially by linearity of expectation. When the payoff function is the max (or min) function, expected utility can be computed via our algorithm above. What if the payoff function is some arbitrary nonlinear convex function? In particular, suppose we have some $f: \mathbb{R}^{n-1} \to \mathbb{R}$, and we have:
$$U_p(s)  = (U_{p1}(z, p, (s_p, s_1)) \dots , U_{pn}(z, p, (s_p, s_n))$$
where we construct a vector including each $U_{pq}$ for all possible $q \neq p$. Then we want to compute:
$$\mathcal{E}(z,p,x) = \mathbb{E}_{s \in x} [f (U_p(s)) ]$$

We first show that there exist choices of $f$ which make this hard. Suppose $f$ is a fixed boolean formula with binary inputs, with each $U_{pq}(z, p, (s_p, s_q)) \in \{0, 1\}$. Then the 3-SAT problem can be reduced to computing whether $\mathcal{E}(z,p,x) \geq 0$ for some arbitrary product distribution $x$ with all nonzero densities, and thus finding this expectation is $NP$-hard. Thus, as long as $P \neq NP$, such a game does not have the polynomial expectation property. 

A further direction here is to continue thinking about generalizing the max-polymatrix algorithm to similar families of functions; for example, what about the family of ``sorted linear" functions, which can be written as a sorting of the inputs, and then a linear combination? For example, the max-function can be seen as sorting in descending order, followed by a linear combination with coefficients $(1, 0, \dots )$. This is easily provable if there are a constant number of leading nonzero terms, but once this can be arbitrary, this is not necessarily generalizable (computing the probability that some value is the $k$th highest in a list requires $\binom{n}{k-1}$ iterations to decide which players have utilities above them).

\subsection{Necessity of Polynomial Expectation}
Part of the spirit of the original question proposed by Papadimitriou and Roughgarden was to address cases of games \textit{without} polynomial expectation property, and argue whether they still have polynomial correlated equilibrium schemes. As it turns out, the example they provided, Max-Polymatrix, does indeed have the polynomial expectation property. This inspires the following general question: Papadimitriou and Roughgarden demonstrate that the polynomial expectation property is a \textit{sufficient} condition for a polynomial correlated equilibrium scheme. To what extent is it \textit{necessary}? 

A small idea in this direction relates to the necessity of the computation of $UX^\top$ in the original algorithm in \cite{papadimitriou2008computing}. In particular, to solve the dual program $[UX^\top]\alpha \geq 0$, it is not entirely necessary to compute $UX^\top$ itself; it suffices to have a separation oracle which, given some input $\alpha$, returns either that it is feasible or a halfspace it violates (via the ellipsoid algorithm). This is equivalent to having an oracle which, given a distribution $X^\top \alpha$\footnote{Note that the entire vector $X^\top \alpha$ might be hard to compute, as we are linearly combining very large vectors. However, we can keep the representation as a linear combination of product distributions and treat ``accessing" any element in this distribution as taking polynomial-time instead of constant.}, returns either that it is a correlated equilibrium or demonstrates a failed constraint. So, we have:
$$\text{Verifying CE in poly-time} \implies \text{Finding CE in poly-time}$$
Having a ``equilibria-checking oracle" is a very similar condition to having polynomial expectation property; in the vast majority of cases, checking a correlated equilibrium requires some computation of expected utility. However, this could address some strange games where computing the exact expected utility is hard, but it is easy to verify the inequalities corresponding to a correlated equilibria.

\bibliographystyle{unsrt}
\bibliography{main}

\begin{thebibliography}{1}

\bibitem{papadimitriou2008computing}
Christos~H Papadimitriou and Tim Roughgarden.
\newblock Computing correlated equilibria in multi-player games.
\newblock {\em Journal of the ACM (JACM)}, 55(3):1--29, 2008.

\end{thebibliography}
\end{document}